# Determination of the nearest-neighbor interaction in VO$_2$ via fractal dimension analysis

Jacob Holder[†], Daniel Kazenwadel[†], Peter Nielaba, Peter Baum*

Universität Konstanz, Fachbereich Physik, 78464 Konstanz, Germany
[†]These authors contributed equally to this work.
*peter.baum@uni-konstanz.de

**The Ising model is one of the simplest and most well-established concepts to simulate phase transformations in complex materials. However, its most central constant, the interaction strength *J* between two nearest neighbors, is hard to obtain. Here we show how this basic constant can be determined with a fractal dimension analysis of measured domain structures. We apply this approach to vanadium dioxide, a strongly correlated material with a first-order insulator-to-metal phase transition with enigmatic properties. We obtain a nearest-neighbor interaction of 13.8 meV, a value close to the thermal energy at room temperature. Consequently, even far below the transition temperature, there are spontaneous local phase-flips from the insulating into the metallic phase. These fluctuations explain several measured anomalies in VO$_2$, in particular the low thermal carrier activation energy and the finite conductivity of the insulating phase. As a method, our fractal dimension analysis links the Ising model to macroscopic material constants for almost any first-order phase transition.**

Phase transitions are a fascinating branch of physics because a wealth of distinctive phenomena can emerge in macroscopic objects from rather simple sets of atomistic interactions. While thermodynamics drives a material into disorder and randomness, the cooperativity between neighboring elements, for example the spins in magnetic materials or adjacent unit cells in crystals, favors self-organization and leads to spontaneous symmetry breaks into intricate domain structures. The Ising model [1] is one of the simplest and most well-established theories [2,3] to understand phase transitions from an atomistic perspective. Basically, multiple discrete elements or cells in an array interact cooperatively with their nearest neighbors while temperature provides a perturbing force. Besides its original use in magnetism [4-6], the Ising model is applied for crystallization and nucleation [7-10], genetics [11] and even social sciences [12]. However, its central constant, the nearest-neighbor interaction *J*, is in most cases not well related to any measurable property of a macroscopic material, and therefore hard to obtain. This lack substantially limits the applicability of the Ising model to predict or understand the properties of a material.

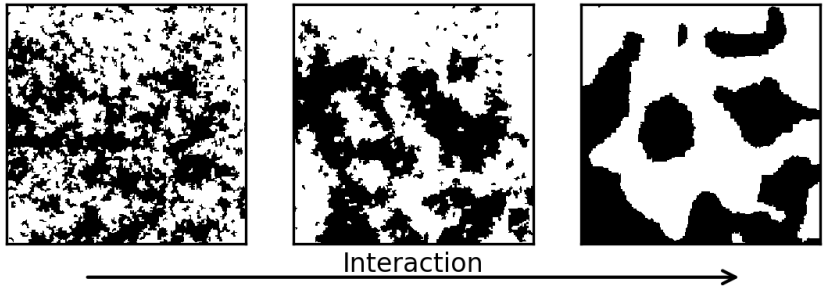

**Fig. 1. Influence of interaction *J* on domain formation.** Left to right, typical domain patterns in a two-phase material for increasing nearest-neighbor interaction *J*. For small *J* (left), neighboring cells hardly interact with each other, favoring a nearly random domain structure and jagged surfaces. For intermediate *J* (middle), intricate domains form via the interplay of random flips from temperature and the cooperativity between adjacent cells. For high *J* (right), the cooperative interactions dominate and produce well-defined domains with smooth surfaces. The ratio between the black and white phases is ~50% in all pictures, but they differ in their fractal geometry.

In this work, we report how to use a fractal dimension analysis of a measured macroscopic domain pattern to determine the underlying nearest-neighbor interaction *J* in a quantitative way on atomistic dimensions. To elucidate the idea of the approach, Fig. 1 depicts a material with two phases (black and white) at a temperature close to a first-order phase transition. At otherwise identical conditions, the interaction strength *J* is increased from left to right, causing the macroscopic domain structure to favor more and more consolidated configurations with reduced surface roughness. This consolidation shows itself as a decrease of the fractal dimension, that is, the scaling behavior of the perimeter-to-area ratio of the domains. The idea of this work is to link, by comparison of an atomistic Ising model with measured domain shapes, the microscopic coupling *J* to a set of measurable quantities from a macroscopic experiment.

We demonstrate our approach on the example of vanadium dioxide ($VO_2$), a strongly correlated material with a notable first-order phase transition from monoclinic/insulator to rutile/metallic at a temperature of $T_t \approx 340$ K [13], slightly above room temperature. This phase transition is not only relevant for technological applications, for example thermochromic windows [14-16], ultrafast photoelectric switches [17,18], ultrasensitive bolometers [19] or impulse strain-wave emitters [20], but it is also of great interest to fundamental physics, because the atomistic transition proceeds on non-trivial reaction paths [21-24] and the strongly correlated nature of the material [25] can bring even advanced ab-initio calculations [26-28] close to the edges of their applicability range. Although VO₂ has been heavily investigated [13-40], there remain many open questions, in particular why insulating VO₂ has several orders of magnitude higher electrical conductivity than estimated from the band gap [29-34], and why there is an unexpected upper limit for it [35,36]. Experiments with Kelvin probe force microscopy (KPFM) [37] and scanning near-field infrared microscopy (SNOM) [38] have revealed a complex set of domain patterns that will be the experimental basis of our report. Similar patterns are observed after laser excitation on ultrafast time scales [39,40].

Let us first consider the general approach of our fractal dimension analysis and discuss the physical implications at a later stage. We consider a material with two phases and describe its microscopic behavior with an Ising model. We use a cubic lattice and identify the two phases (for example metal and insulator) with the two states $\sigma_i = \pm 1$ of the Ising system. Each unit cell of the crystal is associated with one cell in the Ising model. The nearest-neighbor interaction $J$ denotes the energy it takes when two neighboring cells are in different phases. The temperature $T$ in the material is considered via the free energy of a unit cell, approximated as $h(T) = L\frac{T_t-T}{T_t}$, where $T_t$ is the transition temperature. This can be derived using the Helmholtz free energy $F = U - TS$ [41] of a first-order phase transition with the latent heat $L = T\,\Delta S$. The resulting energy bias is zero for cells in any phase at the transition temperature $T_t$ but linearly favors the "correct" phase for cells with a temperature difference $T_t - T$. The resulting Hamiltonian $H$ of our Ising model is

$$H = -\frac{1}{2}\sum_{i,j \in NN} J \cdot \sigma_i \sigma_j + \sum_i h_i(T)\sigma_i \quad (1)$$

where $\sigma_i$ is the state at position $i$ and $\sigma_j$ are all the nearest neighbors at positions $j$. The factor of ½ accounts for the double summation. Note that the energy to create a domain wall in the Ising model is $2J$ because $\sigma_i\sigma_j$ jumps from -1 to 1. All parameters except $J$ are available from experiments; the values for VO₂ are given in Table 1.

| System parameter | Value | Reference |
|---|---|---|
| Transition temperature $T_t$ | 340 K | [13] |
| Standard deviation $\Delta T_t$ | 0.25 K | [38,42] |
| Latent heat $L$ | 51 J/g ≈ 3 $k_B T_t$ | [43] |
| Unit cell size | 0.5 nm | [44] |
| Grain size | 90 nm | [42] |

**Table 1. Parameters in our simulations of VO₂.**

Realistic materials in condensed-matter physics are often not perfect crystals but have inhomogeneities on nanometer and micrometer dimensions, for example, small variations of density, strain or stoichiometry. Such inhomogeneities often slightly change the local transition temperature, and $T_t$ therefore becomes a function of position within the material. Without such deviations, the hysteresis would be infinitely sharp and no domains would be observed. In our Ising model, we consider a finite steepness of the hysteresis curve by assigning a normal-distributed $T_t$ with a width $\Delta T_t$ to each cell in a way that mimics the typical nanostructure of a realistic material. Figure 2a shows a measured scanning-electron-microscopy image of a polycrystalline thin film of $VO_2$ and Fig. 2b depicts the granular $T_t$ map that we use in our simulations (Appendix A). While the existence of a finite bias width is central to our approach, its specific value or the particular distribution into grains does not affect the fractal domain geometry and its link to $J$ (Appendix A).

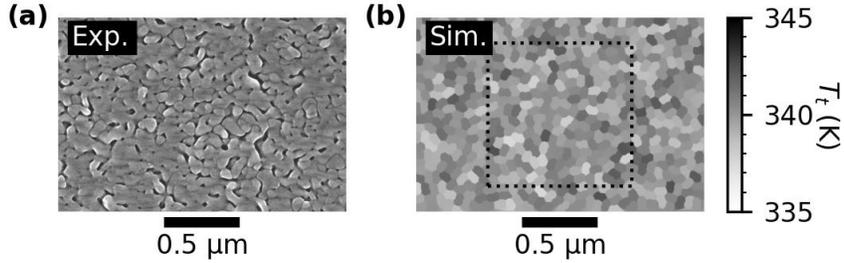

**Fig. 2. Generation of the grain-induced bias of the transition temperature.** (a) Scanning electron microscopy image of a $VO_2$ thin film prepared by sol-gel deposition, reproduced from Ref. [42]. Even after substantial annealing efforts, slightly different grains are still distinguishable. Black to white, secondary electron emission current. (b) Model for the distribution of transition temperature $T_t$ in our simulations (see Appendix A). The dotted lines indicate the area used for the analysis in Fig. 3.

We use the Metropolis Monte Carlo algorithm [45] to generate equilibrated sample states for different interactions and temperatures. We simulate a three-dimensional volume of 2000×2000×120 unit cells (1000×1000×60 nm³) with six nearest neighbors per cell and alternatively a two-dimensional area with 2000×2000 unit cells (1000×1000 nm²) with four nearest neighbors per cell. For each configuration and initialization, we conduct 5000 Monte-Carlo sweeps, found to be sufficient for reaching equilibrium. One typical 3D simulation takes ~640 core hours on the supercomputer JUWELS in Jülich while a typical 2D simulation takes only ~40 core hours on a desktop computer.

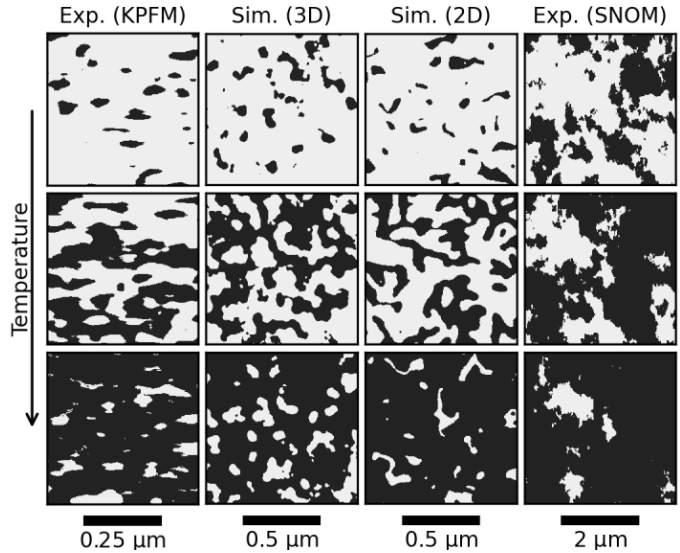

**Fig. 3. Measured and simulated domain structure of VO$_2$ thin films at different temperatures.** Black, metallic phase; white, insulator phase. Top to down, increasing temperature, calibrated by the ratio of phase coverage. KPFM, Kelvin probe force microscopy (left column); SNOM, scanning near-field optical microscopy (right column); data reproduced from Refs. [37,38]. In the 3D simulations, we plot only the topmost layer. Besides slight measurement artifacts, such as line scanning effects in KPFM and non-vanishing depth information in SNOM, the measured and simulated fractal geometries are in good agreement.

Figure 3 shows in the middle two columns a set of simulated domain configurations. In comparison, the left and right columns show measurement results from Kelvin probe force microscopy (KPFM) and scanning near-field optical microscopy (SNOM), reprinted from Refs. [37,38]. From top to down, the temperature is increased. We see that our Ising simulations and the two experimental datasets produce comparable results. The slight discrepancies in the SNOM measurements (right column) are explained by the finite probing depths of ~50 nm in SNOM [46] as compared to <1 nm in KPFM [47].

Sohn et al. [37] have found in their experiments on thin films that the domain patterns have a fractal shape and therefore the ratio of perimeter to area follows a power law. To find the microscopic interaction strength *J* on atomistic dimensions, a metric is needed to compare the simulated domain patterns from the Ising model with the experimental results. Figure 4a depicts our approach: we numerically determine the domain perimeter and domain area for each measured or simulated domain with depth-first search; data is not sorted by temperature in this approach. Figure 4b shows the results for the two experiments (dots and squares). We see that all data points lie on a linear curve in a log-log plot, indicating a power law that is consistent over four orders of magnitude, independently of temperature. The slight constant offset between the two datasets is irrelevant because it relates only to a smoothing of the perimeter by experimental resolution effects.

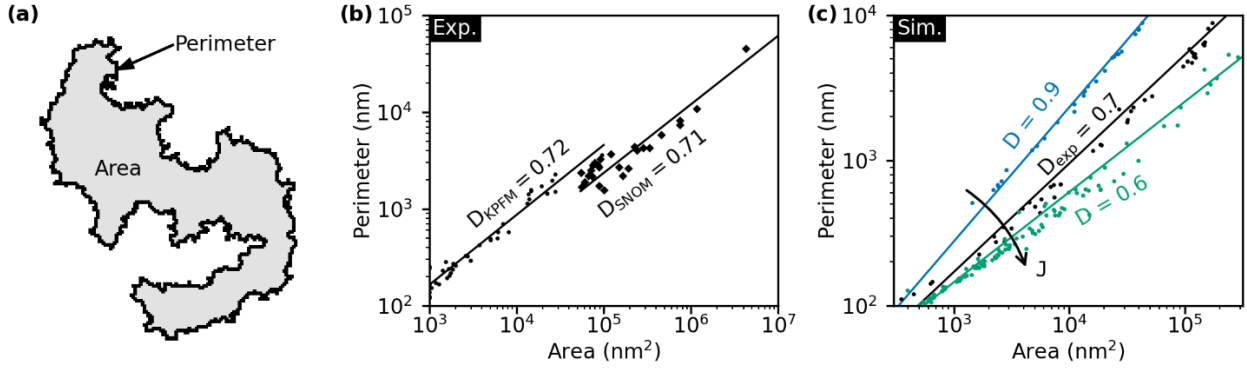

**Fig. 4. Area dependence of the perimeter distribution and analysis of fractal dimensions. (a)** Definition of area and perimeter. **(b)** Experimental data from Kelvin probe force microscopy (dots) and scanning near-field optical microscopy (squares). Self-similarity results in a power law behavior (linear in the log-log plot) in both cases (solid lines). The averaged fractal dimension (the slope) in the two experiments is $D_{exp} = 0.715$. **(c)** Same data from our three-dimensional Ising model for increasing interaction $J$ (blue, black, green) together with the resulting fractal dimensions $D$. The black result corresponds best to the experiment.

In order to determine the fractional exponent, we fit the perimeter-to-area distribution of Fig. 4 by linear regression, weighted by inverse density via Kernel density estimation [48,49] to account for the inhomogeneous size distribution of the domain clusters; there are typically many small ones and few large ones. Finite-size effects [50,51] from the experimental resolution or the discrete nature of the Ising model are avoided by disregarding domains with areas below $10^3$ nm$^2$ for KPFM, below $10^5$ nm$^2$ for SNOM, and below 50 nm$^2$ in the simulations. We define the fractal dimension $D$ as the slope of the linear fits in the log-log plot. It is related to the Hausdorff dimension $D_H$ [52] used in Ref. [37] by a factor of two, that is, $D_H = 2D$; compare Refs. [53-55]. We obtain $D_{KPFM} = 0.72$ and $D_{SNOM} = 0.71$. These almost identical results, obtained with completely unrelated experimental methods, KPFM and SNOM, on rather different thin-film materials, produced with pulsed laser deposition [37] or the sol-gel method [38], show that the fractal dimension of a two-phase material is indeed an intrinsic material constant. The consistency of these two results shows that the atomistic cooperative effects in vanadium dioxide do not depend on sample quality. Therefore, the results of experiments on different kinds of thin films, nanobeams, or bulk crystals can be easily compared.

The fractal dimension of a domain pattern, accessible to experiments, is therefore a robust and solid basis for our next step, the extraction of a value for the microscopic interaction. Figure 4c shows a fractal dimension analysis of simulated domain configurations from our Ising model as a function of an increasing interaction $J$. Using the identical analysis procedure as described above, we find the same power laws as in the experiments. We see that the fractal dimension $D$ of the simulated datasets strongly depends on the atomistic interaction parameter $J$. With decreasing $J$ (blue), the domain perimeters become more and more irregular and jagged, resulting in a higher $D$, while a larger $J$ (green) produces more roundish domains, resulting in a smaller $D$ (compare Fig. 1).

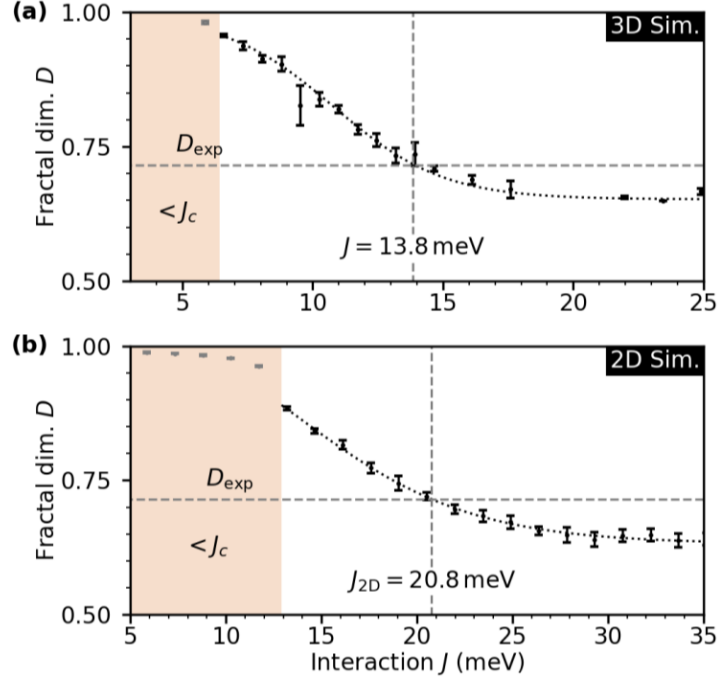

**Fig. 5. Determination of nearest-neighbor interactions $J$ from fractal dimensions $D$. (a)** Results of a set of three-dimensional simulations; we obtain $J \approx 13.8$ meV for the measured $D_{exp}$ (dashed lines). Error bars, standard deviation of five independent simulations. The dotted line is a logistic fit to the data; we exclude data points below the critical interaction $J_C$ where the cells become uncorrelated (orange). **(b)** Results of a set of two-dimensional simulations; we obtain an effective $J_{2D} \approx 20.8$ meV (dashed lines).

Figure 5a shows the dependency between $D$ and $J$ as obtained from the simulations. Each data point is the average of five independent simulations; the error bar denotes the standard deviation. For a very high interaction, the domains become nearly circular and the fractal dimension eventually approaches 1/2. For zero interaction, each cell is completely independent of its neighbors and the fractal dimension approaches 1. Between these limits, we find that the dependency of $D$ on $J$ can be approximated with the logistic function (dotted line). Reproducing the measured fractal geometry $D_{exp} = 0.715$ requires a nearest-neighbor interaction of $J = 13.8$ meV.

In order to describe thin-films or monolayers at much shorter computation times, we alternatively invoke a two-dimensional simulation; the analysis procedure remains the same. Figure 5b shows the results. Again, we see a characteristic increase of the fractal dimensions $D$ with $J$ that allows us to extract a value for the effective two-dimensional interaction; we obtain $J_{2D} = 20.8$ meV. Interestingly, this value relates to the 3D result by a factor of roughly 6/4, the ratio of the numbers of nearest neighbors.

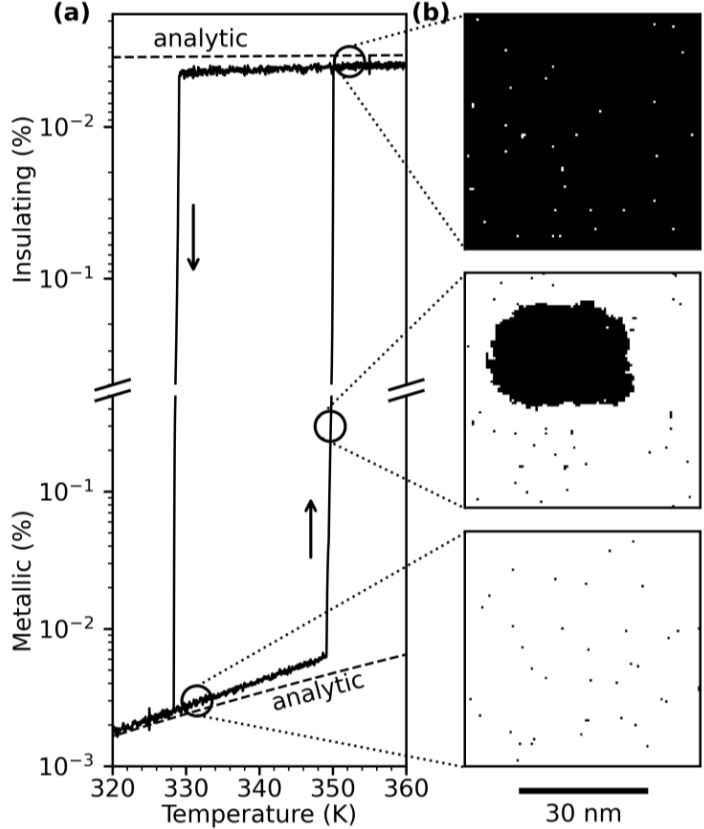

**Fig. 6. Average fraction of metallic respectively insulating phase as a function of temperature. (a)** Fraction of the individual minority phase across the hysteresis curve. When approaching the transition from low temperatures (upwards arrow), there is an increasing amount of metallic unit cells in an otherwise insulating material. The dashed line shows our analytic result from the Hamiltonian of Eq. 1. The slight deviations when approaching the phase jump are due to double-cell or multi-cell fluctuation events. A similar behavior occurs during cooling (downwards arrow). **(b)** Real-space snapshots of the simulated phase maps for three selected temperatures. Bottom, below the transition temperature; middle, after nucleation of a first stable bigger domain; top, far above the transition. Here, the fluctuations are similar but reversed as compared to the low-temperature phase. White denotes insulating/monoclinic unit cells, black denotes metallic/rutile unit cells.

We now discuss the immediate physical consequences of our results on VO$_2$. Close to room temperature, when the material is in principle in the insulating phase, our simulations reveal a substantial percentage of single unit cells that are, for tiny amounts of time, fluctuating into the metallic phase. However, each of these spontaneously flipped unit cells is unstable and rapidly flips back. We can now use the measured $J$ = 13.8 meV to predict quantitatively the fraction of such cells. The activation energy for the creation of a single rutile/metallic unit cell in an otherwise monoclinic/insulating material is $E_{act} \approx 6 \cdot 2J$ = 166 meV. Below the transition, we obtain the average fraction $P_{metal} = e^{-E_{act}/k_B T}$ of metallic unit cells. For example, at room temperature at $T$ = 300 K, about 40 K below $T_t$, where the entire crystal appears completely insulating in the measurements [29], these fluctuations amount to 1‰, averaged over space and time. This level of fluctuations corresponds nicely to the measured carrier density at room temperature from Hall experiments [34,56].

Figure 6 shows the phase distributions between metallic/rutile and insulating/monoclinic at different sample temperatures in the simulations. Figure 6a shows the average fraction of the minority phase as a function of temperature and Fig 6b shows snapshots of the simulated sample area at selected times. As the temperature increases, more and more unit cells flip, thermally activated, into the metallic phase and back (bottom of Fig. 6b). The dashed line shows the expected amount of metallic unit cells as calculated analytically for flips of single unit cells. Far below the transition, it reproduces the simulated behavior. Closer to the transition there appear deviations due to the contributions of fluctuating double-cell and multi-cell nano-domains.

At a certain threshold (here 350 K), the first stable metallic domain forms (middle of Fig. 6b) and rapidly grows, triggering a fast transformation of the complete material (upwards arrow). In the high-temperature phase, the metallic/monoclinic material (top of Fig. 6b) fluctuates back into the insulating/rutile phase (white dots) at a similar rate as in the opposite, insulating case. The slope of the fluctuations is now lower, because a rising temperature now results in higher available thermal energies but also higher flipping barriers (see second part of Eq. 1). When the sample is cooled down, it transforms back into the insulating phase (downwards arrow) with an insulating/monoclinic phase nucleus (white). The width of the simulated hysteresis depends on the speed of the cooling and heating processes. A movie of these fluctuations and the corresponding phase percentages during heating and cooling is provided in the supplementary material.

The existence of these surprisingly substantial thermal phase fluctuations far below the transition temperatures offers an elegant explanation of diverse reported anomalies of VO$_2$. For example, insulating VO$_2$ has an optically measured bandgap of 600 meV [57,58] and the material should therefore be a good insulator with an electrical activation energy of ~300 meV [35,34]. However, many experiments [29-34] report substantially lower values of 90-190 meV, for example, in nanorods (90 meV [29]), nanowires (128 meV [30]), thin films on sapphire (76 meV [31], 168 meV [32]), thin-films on Ge (180 meV [31], 190 meV [33]), thin-films on Si (190 meV [33]), and bulk (100-650 meV [34]). Only experiments on extremely thin films [59] and nanobeams in special geometries [35] report values of 225-310 meV and 300 meV, respectively.

In our calculations, the structural activation energy of rutile/metallic unit cells is $E_{act} \approx 6 \cdot 2J =$ 166 meV. If we assume that a material with fluctuating unit cells has a carrier concentration that is the average of the individual unit cells (mean material), almost all of the experimental results can be reproduced. The steady-state carrier densities in insulating/monoclinic $VO_2$ are a direct result of spontaneous and quick structural unit cell fluctuations into the metallic/rutile phase and back. Band gap effects, special energy levels, doping, defects or edge effects are not required to explain the high conductivity of insulating $VO_2$. The observed decrease of conductivity in strained materials [34,35] fits into the picture, because closer proximity between adjacent unit cells increases the structural coupling $J$ and therefore increases the activation energy for metallic unit cells, reducing their number. In $VO_2$ nanorods with domain coexistence, there is a surprising upper limit for the conductivity of the insulating parts [35,36]. Our model of phase fluctuations naturally explains this result via classical nucleation theory, because only a maximum amount of metallic unit cells can exist within an insulating domain until they nucleate and transform into a macroscopic metallic domain.

In the high-temperature phase, experiments have revealed an increase in electrical conductivity with increasing temperature [33,60,61], opposite to the behavior of normal metals. Our model naturally predicts this effect: Slightly above the transition temperature, there are lots of fluctuating insulating/monoclinic unit cells that act as defects and reduce the electrical conductivity. At higher temperatures above the transition temperature, these fluctuations are more and more suppressed (see Fig. 6a). Consequently, electrical conductivity increases with rising temperature.

The reported spontaneous phase fluctuations affect not only the electronic but also the mechanical properties of $VO_2$: When approaching the phase transition, $VO_2$ becomes soft in terms of an increasing lattice elasticity [62], because the higher amount of structural phase fluctuations at higher temperatures allows the material to react more softly to mechanical strain. A generalized Ising model (Potts model) can be applied if other low-temperature phases, external strain [63], stripe formation [64], twinning or related effects shall be simulated as well.

In summary, our fractal dimension analysis of macroscopic experimental domain structures provides a robust and reproducible material constant, the fractal dimension, that we can relate to the most central parameter of the Ising model, the microscopic cooperativity between adjacent unit cells. This direct access to the interaction strength therefore enables coarse-grained simulations of almost any material with domain structures in a quantitative way. All material parameters that are needed for the reported procedure are quite basic and obtainable for a manifold of other phase-change materials as well. Morphologies or defects that only impact the local transition temperature are insignificant to our results, and fractional dimension analysis therefore gives direct access to the properties of the ideal material. Quantitative knowledge of atomistic nearest-neighbor interactions links the Ising model to macroscopic quantities of a material, such as the absolute energy cost for the formation of domain boundaries, and also provides insight into fluctuation effects. These abilities will help to elucidate the emergence of macroscopic phase diagrams from microscopic phenomena and link the dynamics of phase transitions to atomistic processes in space and time.

**Appendix A: Modelling of crystal nanostructures.** For simulating inhomogeneous crystals with varying $T_t$ (see Fig. 2a); we use a 90% volume fraction of dumbbell colloids with an aspect ratio of 1.55 [65] to represent the different grains. The dumbbells have a length of 90 nm [42]. Each grain gets a normally distributed transition temperature around $T_t$ with a width of $\Delta T_t = 0.25$ K, estimated from measured hysteresis curves [38,42]. We close the remaining holes in the $T_t$ map by cubic interpolation and ensure periodic boundaries by stitching of mirror images. Figure 1b shows the results. Variations of $\Delta T_t$ or different grain sizes do not affect our simulations, even when the grain size becomes as small as one unit cell. Although perfect bulk single-crystals cannot be used for our procedure because the domains are as large as the entire material, the robustness of our results over 3-4 orders of magnitude of domain sizes (see Fig. 4b) indicates the applicability of our evaluated $J$ for single crystals on millimeter dimensions [66]. Materials in which anisotropic external strain is dominant [63,64], could be incorporated into our model by a generalized Ising Hamiltonian.

**Acknowledgments:** We thank A. Sohn and D.-W. Kim for sharing the data reproduced in Fig. 4, and A. Lüders for providing dumbbell shapes. We acknowledge financial support by the Evangelisches Studienwerk e.V. and the Deutsche Forschungsgemeinschaft through Sonderforschungsbereich SFB 1432. We gratefully acknowledge the computing time granted by the John von Neumann Institute for Computing (NIC) on JUWELS at the Jülich Supercomputing Centre (JSC).